\documentclass[11pt,a4paper]{article}
\usepackage{amsfonts,amssymb,epsfig,amsmath,enumitem}
\usepackage[utf8]{inputenc}
\usepackage{textcomp}
\usepackage{caption,subcaption}
\usepackage{xcolor}
\usepackage{tikz}
\usepackage{hyperref}

\renewcommand{\thefootnote}{\fnsymbol{footnote}}

\begin{document}

\makeatletter \@addtoreset{equation}{section} \makeatother
\renewcommand{\theequation}{\thesection.\arabic{equation}}
\renewcommand{\thefootnote}{\alph{footnote}}

\begin{titlepage}
\begin{center}

\hfill {\tt}\\

\vspace{2cm}

{\Large\bf Testing 5d-6d dualities with fractional D-branes} 

\vspace{2cm}

{\large Youngbin Yun$^1$}

\vspace{0.7cm}

\textit{$^1$Department of Physics and Astronomy \& Center for
Theoretical Physics,\\
Seoul National University, Seoul 151-747, Korea.}\\

\vspace{0.7cm}

E-mails: {\tt ybyun90@gmail.com}

\end{center}

\vspace{1cm}

\begin{abstract}

6d SCFTs compactified on a circle can often be studied from nonperturbative 5d super-Yang-Mills theories, using instanton solitons. 
However, the 5d Yang-Mills theories with 6d UV fixed points frequently have too many hypermultiplet matters, which makes it difficult to use 
the ADHM techniques for instantons.  With the examples of 6d $\mathcal{N}=(1,0)$ SCFTs with $Sp(N)$ gauge symmetry
and $2N+8$ fundamental hypermultiplets, we show that one can still make rigorous studies of these 5d-6d relations in 
the `fractional D-brane sectors'. We test the recently proposed 5d duals given by $Sp(N+1)$ gauge theories, and compare their 
instanton partition functions with the elliptic genera of 6d self-dual strings.

\end{abstract}

\end{titlepage}

\tableofcontents

\section{Introduction} 
\label{sec:intro}
In this paper we study the recently conjectured 5d gauge theory descriptions of 6d SCFTs compactified on a circle \cite{Hayashi:2015zka}. The 6d $\mathcal{N}=(1,0)$ SCFTs have a tensor multiplet and $Sp(N)$ gauge symmetry with $N_f=2N+8$ fundamental hypermulitplets, and these can be Higgsed \cite{Kim:2015fxa} to the E-string theory \cite{Ganor:1996mu,Klemm:1996hh,Minahan:1998vr,Eguchi:2002fc}. The 5d $\mathcal{N}=1$ gauge theories have $Sp(N+1)$ gauge symmetry and $N_f=2N+8$ fundamental hypermultiplets. The 5d $Sp(N+1)$ gauge theories at $N\geq1$ with $N_f \leq 2N+6$ hypermultiplets are known to have non-trivial 5d UV fixed points \cite{Seiberg:1996bd,Intriligator:1997pq}. If $N_f \geq 2N+7$, the theories have Landau pole issues, in that the Coulomb branch moduli spaces are incomplete by having strong coupling singularities. So for such theories to have UV fixed points, there should be physical explanations of these singularities. Since the 5d descriptions suggested in \cite{Hayashi:2015zka} have $N_f=2N+8$ matters beyond the bound of \cite{Seiberg:1996bd,Intriligator:1997pq}, it would be desirable to have a better understanding on how this is happening. \\

In this paper we test these novel 5d-6d dualities by studying the spectrum of instanton solitons. Instantons in the 5d gauge theories play important roles in studying 5d and 6d SCFTs \cite{Tachikawa:2015mha,Yonekura:2015ksa,Kim:2015jba}. In particular, for 5d gauge theories having 6d UV fixed points, instantons are Kaluza-Klein momenta on the compactified circle. Therefore instantons are crucial objects in 5d to understand the 6d physics.\\

To study instantons, we use the ADHM construction engineered by string theory brane picture. However, since the ADHM construction embeds the instanton quantum mechanics into string theory, it often contains unwanted extra degrees of freedom which are not included in the QFT that one is interested in. So when we compute instanton partition functions via the string theory engineered ADHM partition functions, the extra contribution part should be subtracted to obtain correct QFT instanton partition functions \cite{Hwang:2014uwa}. For various models one could separately compute these extra contributions from string theory considerations \cite{Hwang:2014uwa}.  Unfortunatley we don't know how to compute these extra contributions for the 5d $Sp(N+1)$ gauge theories with $N_f=2N+8$ hypermultiplets. Nonetheless one can compute one-instaton function exactly. Brane system for the 5d gauge theories have an $O7^-$-plane, and one-instanton sector is described by the half-D1-brane localized at the $O7^-$-plane. One expect that there is no extra degrees of freedom in one-instanton sector. See section 2.2 and Figure~\ref{5dbrane}. \\

Instanton partition functions compute the BPS spectrum of instantons bound to W-bosons in the Coulomb phase. Part of 5d W-bosons uplift to 6d self-dual strings wrapping the circle, and instantons are KK momenta on these strings. So one can study the same physics from the elliptic genera of 6d self-dual strings. We compute these elliptic genera, and compared them with the one-instanton partition function of the 5d gauge theories. We find perfect agreements, which provide nontrival supports of the proposal made in \cite{Hayashi:2015zka}. In particular, our test clarifies the physical setting of the 5d-6d dualities, by emphasizing the roles of background Wilson lines,
and also by explicitly showing the relations between various 5d and 6d parameters.\\

This paper is organized as follows. In section 2, we will briefly review the E-string theory and $Sp(N)$ generalizations. In both cases, it is crucial to consider the effects of background Wilson lines for the flavor symmetries. We compare the E-strings' elliptic genera and the 5d instanton partition functions combined with perturbative index, and show the fugacity map of the two indices. In section 3, we compute elliptic genera for self-dual strings in the 6d SCFT with $Sp(1)$ gauge symmetry and 10 hypermultiplets using 2d gauge theory description. We compare this result with the one-instanton partition functions of the 5d $Sp(2)$ gauge theory with 10 fundamental hypermultiplets. In section 4, we will generalize our result to 6d $Sp(N)$ gauge theories. We can see that 5d $Sp(N+1)$ gauge group can be decomposed into the $Sp(1)\times Sp(N)$, and the former $Sp(1)$ gives the 6d self-string structure as E-string theory and the latter $Sp(N)$ gives 6d gauge group. In section 5, we will conclude with some remarks on the future direction.\\

\section{E-strings and their $Sp(N)$ generalizations}
\label{sec:review}
We will briefly review the E-string theory \cite{Ganor:1996mu,Klemm:1996hh,Minahan:1998vr,Eguchi:2002fc,Kim:2014dza}, and their circle compactifications to the 5d $Sp(1)$ gauge theory with 8 hypermultiplets. The E-string theory and 6d $\mathcal{N}=(1,0)$ SCFT with $Sp(1)$ gauge symmetry are well-studied in reference \cite{Kim:2014dza,Kim:2015fxa}, and we will follow their idea. \\

\begin{figure}
\centering
\begin{tikzpicture} 

\draw [blue,thick] (-0.05,-2) -- (-0.05,2);
\draw [dashed] (0,-2) -- (0,2);
\draw [blue,thick] (0.05,-2) -- (0.05,2);
\draw [thick] (-3.3,0.05) -- (3.3,0.05);
\draw [thick] (-3.3,0) -- (3.3,0);
\draw [thick] (-3.3,-0.05) -- (3.3,-0.05);
\draw [thick,red] (0,0.12) -- (2.5,0.12);
\draw [thick,red] (0,0.17) -- (2.5,0.17);
\filldraw [black!40,draw=black!100,thick] (2.5,0) circle (0.25cm);
\filldraw [black!40,draw=black!100,thick] (-2.5,0) circle (0.25cm);
\node  at (1.1,-0.5) {$2N$ D6s};
\node  at (1.1,0.5) [red] {$n$ D2s};
\node  at (2.5,-0.5) [black] {NS5};
\node  at (1.8,1.8) [blue] {O8$^-$ - 8 D8s};

\draw (6,2) --(6,-1.2);
\draw (4.3,1.3) --(12.3,1.3);

\node  at (5.2,0.9) {D2};
\node  at (5.2, 0.3) {NS5};
\node  at (5.2, -0.3) {D6};
\node  at (5.2,-0.9) {O8$^-$-D8};

\node at (6.5,1.7) {0}; \node at (7.1,1.7) {1}; \node at (7.7,1.7) {2}; \node at (8.3,1.7) {3}; \node at (8.9,1.7) {4}; \node at (9.5,1.7) {5}; \node at (10.1,1.7) {6}; \node at (10.7,1.7) {7}; \node at (11.3,1.7) {8}; \node at (11.9,1.7) {9}; 
\node at (6.5,0.9) {$\bullet$}; \node at (7.1,0.9) {$\bullet$}; \node at (7.7,0.9) {-}; \node at (8.3,0.9) {-}; \node at (8.9,0.9) {-}; \node at (9.5,0.9) {-}; \node at (10.1,0.9) {$\bullet$}; \node at (10.7,0.9) {-}; \node at (11.3,0.9) {-}; \node at (11.9,0.9) {-}; 
\node at (6.5,0.3) {$\bullet$}; \node at (7.1,0.3) {$\bullet$}; \node at (7.7,0.3) {$\bullet$}; \node at (8.3,0.3) {$\bullet$}; \node at (8.9,0.3) {$\bullet$}; \node at (9.5,0.3) {$\bullet$}; \node at (10.1,0.3) {-}; \node at (10.7,0.3) {-}; \node at (11.3,0.3) {-}; \node at (11.9,0.3) {-};
\node at (6.5,-0.3) {$\bullet$}; \node at (7.1,-0.3) {$\bullet$}; \node at (7.7,-0.3) {$\bullet$}; \node at (8.3,-0.3) {$\bullet$}; \node at (8.9,-0.3) {$\bullet$}; \node at (9.5,-0.3) {$\bullet$}; \node at (10.1,-0.3) {$\bullet$}; \node at (10.7,-0.3) {-}; \node at (11.3,-0.3) {-}; \node at (11.9,-0.3) {-};
\node at (6.5,-0.9) {$\bullet$}; \node at (7.1,-0.9) {$\bullet$}; \node at (7.7,-0.9) {$\bullet$}; \node at (8.3,-0.9) {$\bullet$}; \node at (8.9,-0.9) {$\bullet$}; \node at (9.5,-0.9) {$\bullet$}; \node at (10.1,-0.9) {-}; \node at (10.7,-0.9) {$\bullet$}; \node at (11.3,-0.9) {$\bullet$}; \node at (11.9,-0.9) {$\bullet$};

\end{tikzpicture}
\caption{type IIA brane system for 6d $\mathcal{N}=(1,0)$ $Sp(N)$ gauge theory with $N_f=2N+8$ fundamental hypermultiplets. $n$ D2 branes engineer $n$ self-dual strings. }
\label{6dbrane}
\end{figure}
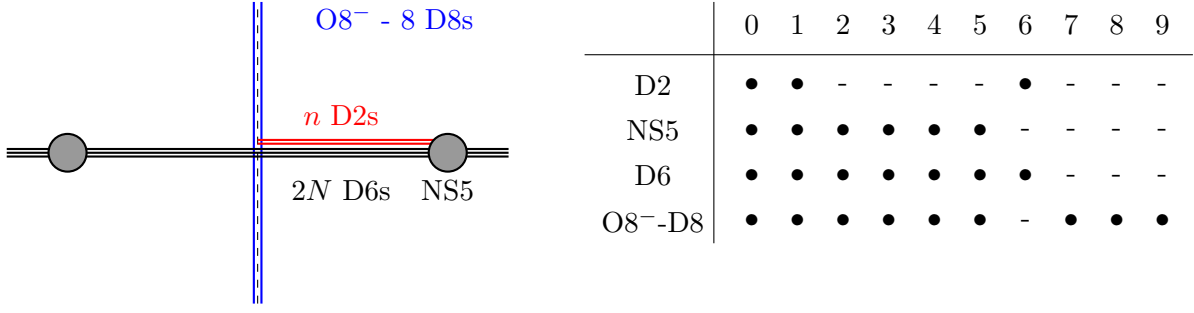

\begin{figure}
\centering
\begin{tikzpicture} 
\draw [thick] (-3.1,-2) circle (0.6cm);
\filldraw [white!100!,draw=black!100,thick] (-2,-2) circle (0.8cm);
\draw [thick] (1,-2.8) rectangle (3.5,-1.2);
\draw [thick] (-2.8,-0.2) rectangle (-1.2,0.8);
\draw [thick,dashed] (-1.2,-2) -- (1,-2);
\draw [thick] (-2,-1.2) -- (-2, -0.2);
\node  at (-2,-2)  {$O(n)$};
\node  at (-2.0,0.3)  {$Sp(N)$};
\node  at (2.25,-2)  {$SO(4N+16)$};
\node  at (-4.2,-2.4)  {sym.};
\end{tikzpicture}
\caption{2d ADHM quiver diagram for the self-dual strings}
\label{2dquiver}
\end{figure}
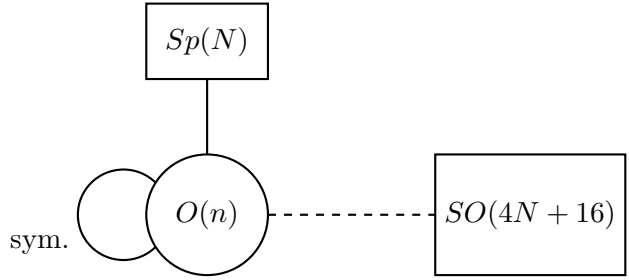

First consider type IIA brane description of the 6d $\mathcal{N}=(1,0)$ SCFT with $Sp(N)$ gauge symmetry and $N_f=2N+8$ hypermultiplets. The case with $N=0$ engineers the E-string theory. 
Brane system is given in Figure~\ref{6dbrane} \cite{Brunner:1997gf,Hanany:1997gh}, and this theory is also known as $(D_{N+4},D_{N+4})$ minimal conformal matter theory \cite{Heckman:2013pva,DelZotto:2014hpa,Heckman:2014qba,Heckman:2015bfa}. We focus on the self dual-strings which couple to the tensor multiplet in the 6d SCFT. The self-dual strings are instanton soliton strings in 6d gauge theory, and it is realized as D2-branes living on D6-branes. The quiver diagram for the 2d $\mathcal{N}=(0,4)$ gauge theory living on D2-branes is given in Figure~\ref{2dquiver}. Their SUSY and Lagrangian are studied in \cite{Kim:2014dza,Kim:2015fxa}.
$O(n)$ vector multiplet and symmetric hypermultiplet come from the strings stretch between D2-D2 branes with appropriate boundary conditions in the presence of O8$^{-}$-plane. Hypermultiplets whose representation is $(n,2N)$ come from D2-D6 strings, and Fermi multiplets whose representation is $(n,4N+16)$ come from D2-D8 strings and D2-D6 strings across NS5 brane. We circle compactify the theory along $x^1$ direction.

\subsection{The elliptic genera of self-dual strings}
We focus on elliptic genera of the self-dual strings of the 6d $Sp(N)$ theories
\begin{align}
Z^{\textrm{6d},Sp(N)} = 1 + \sum_{n=1}^{\infty} w^n Z_n^{\textrm{6d},Sp(N)} \,.
\end{align}
where $w$ is the fugacity for the string winding number.
The elliptic genus of the 2d gauge theory on a tours is
\begin{align}
Z_n^{\textrm{6d},Sp(N)} = \textrm{Tr}_{RR} \left[ (-1)^F q^{2H_L} \bar{q}^{2H_R} e^{2\pi i \epsilon_1(J_1+J_R)} e^{2\pi i \epsilon_1(J_2+J_R)} 
\prod_{i=1}^{N} e^{2\pi i \alpha_i G_i} \prod_{l=1}^{N_f=2N+8}e^{2\pi i m_l F_l} \right] \,.
\label{ellipticg}
\end{align}
$q \equiv e^{i \pi \tau}$ contains the complex structure of the torus $\tau$.\footnote{We use definition of $q$ as $q \equiv e^{i \pi \tau}$ instead of usual $q \equiv e^{2i \pi \tau}$, because instanton fugacity in 5d gauge theory correspond with this definition of $q$.} $H_R \sim \{ Q , Q^{\dagger}\}$ where $Q,Q^{\dagger}$ are $(0,2)$ supercharges of the theory. $J_1,J_2$ and $J_R$ are Cartans of $SO(4)_{2345}$ and $SO(3)_{789} \sim SU(2)_R$. $G_i$ are Cartans of $Sp(N)$ gauge group of 6d SCFT and $\alpha_i$ are corresponding chemical potentials. $F_l$ are Cartans of $SO(4N+16)$ flavor symmetry and $m_l$ are corresponding chemical potentials. The elliptic genus of  $n$ E-strings is given by $Z_n^{\textrm{E-strings}} \equiv Z_n^{\textrm{6d},Sp(0)}$. \\

The elliptic genus of the 2d gauge theory \eqref{ellipticg} was studied in \cite{Benini:2013nda,Benini:2013xpa,Gadde:2013ftv}, and the E-string case(or $O(n)$ gauge group) was further studied in \cite{Kim:2014dza,Kim:2015fxa}. The elliptic genus is given by an integral over the $O(n)$ flat connections on $T^2$.
$O(n)$ gauge group has two disconnected parts $O(n)^{\pm}$. So the Wilson lines $U_1$, $U_2$ along the temporal and spatial circle have two disconnected sector. The discrete holonomy sectors for $O(n)$ gauge group on $T^2$ are listed in section 3 of \cite{Kim:2014dza}. Usually elliptic genus is given by sum of 8 discrete sectors for a given $n$. But $n=1$ and $n=2$ cases are special, and they are given by sum of 4 and 7 sectors respectively.\\ 

The elliptic genus \eqref{ellipticg} is given by \cite{Benini:2013xpa,Kim:2014dza}
\begin{align}
Z_n^{\textrm{6d},Sp(N)} = \sum_{I} \frac{1}{|W_I|} \frac{1}{(2\pi i)^r} \oint Z_{\textrm{1-loop}}^{(I)} \; \,, \quad  Z_{\textrm{1-loop}}^{(I)}  \equiv Z^{(I)}_{\textrm{vec}}Z^{(I)}_{\textrm{sym.}}Z^{(I)}_{\textrm{Fermi}}Z^{(I)}_{\textrm{fund.}} \,.
\end{align}
The 1-loop determinant for the 2d multiplets are given by
\begin{align}
Z_{\textrm{vec}} \;&=\; \prod_{i=1}^{r} \left( \frac{2\pi \eta^2 du_i}{i} \cdot \frac{\theta_1(2\epsilon_+)}{i \eta}  \right) \prod_{\alpha \in \textrm{root}} \frac{\theta_1(\alpha(u)) \theta_1(2\epsilon_+ + \alpha(u))}{i\eta^2} \,, \\
Z_{\textrm{sym hyper}} \;&=\; \prod_{\rho \in \textrm{sym}} \frac{i \eta}{\theta_1(\epsilon_1 + \rho(u))} \frac{i \eta}{\theta_1(\epsilon_2 + \rho(u))} \,, \\
Z^{SO(4N+16)}_{\textrm{Fermi}} \;&=\; \prod_{\rho \in \textrm{fund}} \prod_{l=1}^{2N+8}\frac{\theta_1(m_l + \rho(u))}{i\eta} \,, \\
Z^{Sp(N)}_{\textrm{fund hyper}} \;&=\; \prod_{\rho \in \textrm{fund}} \prod_{i=1}^{N} \frac{i \eta}{ \theta_1(\epsilon_+ + \rho(u) + \alpha_i) }\frac{i \eta}{ \theta_1(\epsilon_+ + \rho(u) - \alpha_i) }	 \,,
\end{align}
where $\epsilon_{\pm} \equiv \frac{\epsilon_1\pm\epsilon_2}{2}$ and $r$ is the rank of  the gauge group $O(n)$. $\eta \equiv \eta(\tau)$ is the Dedekind eta function and $\theta_i(z) \equiv \theta_i(\tau,z)$ are the Jacobi theta functions.
`$I$' refers the disconnected holonomy sectors and $u_i$ are zero modes of 2d gauge fields along the torus. $|W_I|$ is order of Weyl group of $O(n)_I$ for each sector `$I$' \cite{Kim:2014dza}.
For later convenience, we will use following fugacity notation $t \equiv e^{2\pi i \epsilon_+}\,, \; u \equiv e^{2\pi i \epsilon_-}\,,\;v_i \equiv e^{2\pi i \alpha_i}\,, \; y_l\equiv e^{2\pi i m_l}$. The elliptic genus contains contour integral of $u_i$, which is a residue sum given by Jeffrey-Kirwan residue(JK-residue) prescription \cite{Benini:2013nda,Benini:2013xpa}.\\

The E-string elliptic genus has manifest $E_8$ global symmetry. One should turn on the $E_8$ Wilson line on a circle to obtain 5d SYM description of E-string theory \cite{Kim:2014dza}.\footnote{This shift can be naturally understood by embedding the 6d SCFT into M-theory. Namely, to obtain the D4-D8-08 which realizes 5d SYM description, one has to compactify the M5-M9 system on a circle with a Wilson line that breaks $E_8$ to $SO(16)$.} This background $E_8$ Wilson line provides the extra shift $m_8 \rightarrow m_8 - \tau$ to the chemical potential. So it gives following shift of the theta functions
\begin{align}
\theta_i(m_8) \rightarrow \pm\left( \frac{y_8}{q} \right) \theta_i(m_8) \,, 
\end{align}
where we have $(-)$ sign for $i=1,4$ and $(+)$ sign for $i=2,3$. The overall factor shifts by $\frac{y_8}{q}$ can be absorbed by the redefinition of the string winding fugacity $w \rightarrow wqy_8^{-1}$ \cite{Kim:2014dza}. We shall observe later that the $E_8$ Wilson line effect continues to be crucial for the 6d $Sp(N)$ generalizations of the E-string theory.

\paragraph{One-string} With the effect of $E_8$ Wilson line, one-string elliptic genus is given by the sum of 4 discrete sectors
\begin{align}
Z_{n=1}^{\textrm{E-string}}
&= \frac{1}{2}\left( -Z_{1,[1]} +Z_{1,[2]} +Z_{1,[3]} -Z_{1,[4]} \right) \,,
\end{align}
where $Z_{1,[I]}$ for $I=1,2,3,4$ are given by
\begin{align}
Z_{1,[I]} = 
 -\frac{\eta^2}{\theta_1(\epsilon_1) \theta_1(\epsilon_2)} \prod_{l=1}^{8} \frac{\theta_I(m_l)}{\eta} \,.
\end{align}
In order to compare this result with the 5d instanton partition functions, we expand this result in terms of $q$
\begin{align}
Z^{\textrm{E-string}}_{n=1} &= \frac{t}{(1-tu)(1-t/u)}\chi^{SO(16)}_{16}(y_i)q^0 + \frac{t}{(1-tu)(1-t/u)}\chi^{SO(16)}_{\overline{128}}(y_i)q^1 +\mathcal{O}(q^2) \,,
\end{align}
where $\chi^{SO(16)}_{\textrm{R}}$ denotes $SO(16)$ character of representation R.
\paragraph{Two-strings} At n=2, the elliptic genus is given by the sum of 7 sectors. We skip the details of the calculation here, because we shall see the calculation with $Sp(N)$ generalizations in Section \ref{sec:rank1}. We just report the q-expanded two-strings result in the presence of $E_8$ Wilson line
\begin{align}
Z^{\textrm{E-string}}_{n=2} 
& = \left( -\frac{t\;(t+\frac{1}{t})}{(1-tu)(1-t/u)}+\frac{1}{2}\left( \left(\frac{t\;\chi^{SO(16)}_{16}(y_i)}{(1-tu)(1-t/u)}\right)^2+\left(\frac{t^2\;\chi^{SO(16)}_{16}(y_i^2)}{(1-t^2u^2)(1-t^2/u^2)}\ \right)\right)\right)q^0 \\
& \quad + \left( (\frac{t}{(1-tu)(1-t/u)})^2\chi^{SO(16)}_{16}(y_i)\chi^{SO(16)}_{\overline{128}}(y_i) - \frac{t\; (t+\frac{1}{t})}{(1-tu)(1-t/u)}\chi^{SO(16)}_{128}(y_i) \right) q +\mathcal{O}(q^2) \,.
\end{align}

\subsection{5d SYM and instanton partition functions}

\begin{figure}
\centering
\begin{tikzpicture} 

\draw [thick] (-0.1,-0.1) -- (0.1,0.1);
\draw [thick] (0.1,-0.1) -- (-0.1,0.1);
\filldraw [black!40,draw=black!100,thick] (0.3,0.2) circle (0.07cm);
\filldraw [black!40,draw=black!100,thick] (0.3,-0.2) circle (0.07cm);
\filldraw [black!40,draw=black!100,thick] (-0.3,0.2) circle (0.07cm);
\filldraw [black!40,draw=black!100,thick] (-0.3,-0.2) circle (0.07cm);

\draw [thick] (-1.3,3.05) -- (-2.3,2.05) -- (-2.3,1.75) -- (-2.0,1.45) -- (-1.5,1.2) -- (-1,0.7) -- (1,0.7) -- (1.5,1.2) -- (2.0,1.45) -- (2.3,1.75) -- (2.3,2.05) -- (1.3,3.05);
\draw [thick] (-1.3,-3.05) -- (-2.3,-2.05) -- (-2.3,-1.75) -- (-2.0,-1.45) -- (-1.5,-1.2) -- (-1,-0.7) -- (1,-0.7) -- (1.5,-1.2) -- (2.0,-1.45) -- (2.3,-1.75) -- (2.3,-2.05) -- (1.3,-3.05);

\filldraw  [blue!60,draw=blue!60,thick]  (-1.3,3.05) circle (0.07cm);
\filldraw  [blue!60,draw=blue!60,thick]  (-1.3,-3.05) circle (0.07cm);
\filldraw  [blue!60,draw=blue!60,thick]  (1.3,3.05) circle (0.07cm);
\filldraw  [blue!60,draw=blue!60,thick]  (1.3,-3.05) circle (0.07cm);

\draw [thick] (-1,0.7) -- (-1,-0.7) ;
\draw [thick] (1,0.7) -- (1,-0.7) ;

\draw [thick] (1+0.5,0.7+0.5)--(-1-0.5,0.7+0.5);
\draw [thick] (1+0.5,-0.7-0.5)--(-1-0.5,-0.7-0.5);

\draw [thick] (2.0,1.45) -- (3.0,1.45);
\draw [thick] (-2.0,1.45) -- (-3.0,1.45) ;
\draw [thick] (2.0,-1.45) -- (3.0,-1.45) ;
\draw [thick]  (-2.0,-1.45) -- (-3.0,-1.45) ;

\draw [thick] (2.3,1.75) -- (3,1.75) ;
\draw [thick] (-2.3,1.75) -- (-3,1.75) ;
\draw [thick] (2.3,-1.75) -- (3,-1.75) ;
\draw [thick] (-2.3,-1.75) -- (-3,-1.75) ;

\draw [thick] (2.3,2.05) -- (3,2.05) ;
\draw [thick] (-2.3,2.05) -- (-3,2.05) ;
\draw [thick] (2.3,-2.05) -- (3,-2.05) ;
\draw [thick] (-2.3,-2.05) -- (-3,-2.05) ;

\draw [thick,,red!100] (-1,0) -- (1,0) ;
\draw [thick,,red!100] (-1.65,2.7) -- (1.65,2.7) ;
\draw [thick,,red!100] (-1.65,-2.7) -- (1.65,-2.7) ;

\node [red!100] at (0.9,0.3) {$\frac{1}{2}$ D1};
\node [red!100] at (0.9,3) {D1};
\node at (0,-0.5) {O7-4 D7};

\draw [thick] [->] [red!100] (0,2.8) --(0,3.2); 
\draw [thick] [->] [red!100] (0,-2.8) --(0,-3.2); 

\draw (6,2) -- (6,-1.2);
\draw (4.3,1.3) --(12.3,1.3);

\node  at (5.2,0.9) {D1};
\node  at (5.2, 0.3) {NS5};
\node  at (5.2, -0.3) {D5};
\node  at (5.2,-0.9) {O7-D7};

\node at (6.5,1.7) {0}; \node at (7.1,1.7) {1}; \node at (7.7,1.7) {2}; \node at (8.3,1.7) {3}; \node at (8.9,1.7) {4}; \node at (9.5,1.7) {5}; \node at (10.1,1.7) {6}; \node at (10.7,1.7) {7}; \node at (11.3,1.7) {8}; \node at (11.9,1.7) {9}; 
\node at (6.5,0.9) {$\bullet$}; \node at (7.1,0.9) {-}; \node at (7.7,0.9) {-}; \node at (8.3,0.9) {-}; \node at (8.9,0.9) {-}; \node at (9.5,0.9) {-}; \node at (10.1,0.9) {$\bullet$}; \node at (10.7,0.9) {-}; \node at (11.3,0.9) {-}; \node at (11.9,0.9) {-}; 
\node at (6.5,0.3) {$\bullet$}; \node at (7.1,0.3) {$\bullet$}; \node at (7.7,0.3) {$\bullet$}; \node at (8.3,0.3) {$\bullet$}; \node at (8.9,0.3) {$\bullet$}; \node at (9.5,0.3) {$\bullet$}; \node at (10.1,0.3) {-}; \node at (10.7,0.3) {-}; \node at (11.3,0.3) {-}; \node at (11.9,0.3) {-};
\node at (6.5,-0.3) {$\bullet$}; \node at (7.1,-0.3) {$\bullet$}; \node at (7.7,-0.3) {$\bullet$}; \node at (8.3,-0.3) {$\bullet$}; \node at (8.9,-0.3) {$\bullet$}; \node at (9.5,-0.3) {-}; \node at (10.1,-0.3) {$\bullet$}; \node at (10.7,-0.3) {-}; \node at (11.3,-0.3) {-}; \node at (11.9,-0.3) {-};
\node at (6.5,-0.9) {$\bullet$}; \node at (7.1,-0.9) {$\bullet$}; \node at (7.7,-0.9) {$\bullet$}; \node at (8.3,-0.9) {$\bullet$}; \node at (8.9,-0.9) {$\bullet$}; \node at (9.5,-0.9) {-}; \node at (10.1,-0.9) {-}; \node at (10.7,-0.9) {$\bullet$}; \node at (11.3,-0.9) {$\bullet$}; \node at (11.9,-0.9) {$\bullet$};

\end{tikzpicture}
\caption{type IIB brane diagram for the 5d $\mathcal{N}=1$ $Sp(2)$ gauge theory with $N_f=10$ hypermultiplets. The figure shows the covering space of $\mathbb{Z}_2$ quotient by O7 (the cross in the figure). The blue dots denote 7-branes on which vertical 5-branes can end. Half-D1 brane is stuck to the O7$^-$-plane. }
\label{5dbrane}
\end{figure}
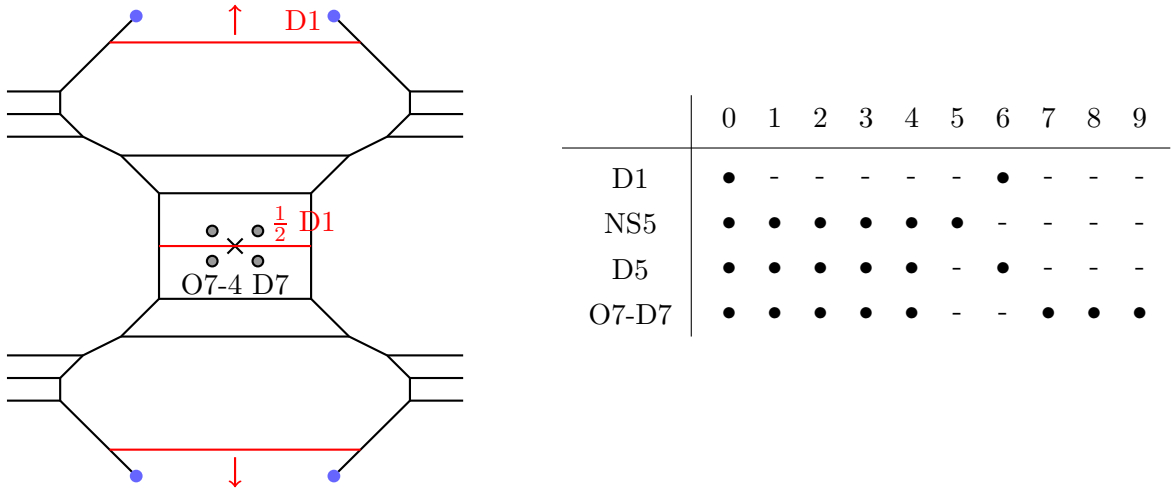

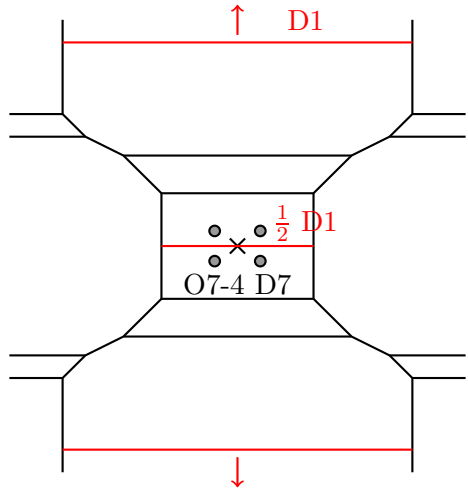
\begin{figure}
\centering
\begin{tikzpicture} 

\draw [thick] (-0.1,-0.1) -- (0.1,0.1);
\draw [thick] (0.1,-0.1) -- (-0.1,0.1);
\filldraw [black!40,draw=black!100,thick] (0.3,0.2) circle (0.07cm);
\filldraw [black!40,draw=black!100,thick] (0.3,-0.2) circle (0.07cm);
\filldraw [black!40,draw=black!100,thick] (-0.3,0.2) circle (0.07cm);
\filldraw [black!40,draw=black!100,thick] (-0.3,-0.2) circle (0.07cm);

\draw [thick]  (-2.3,3) -- (-2.3,1.75) -- (-2.0,1.45) -- (-1.5,1.2) -- (-1,0.7) -- (1,0.7) -- (1.5,1.2) -- (2.0,1.45) -- (2.3,1.75) -- (2.3,3) ;
\draw [thick]  (-2.3,-3) -- (-2.3,-1.75) -- (-2.0,-1.45) -- (-1.5,-1.2) -- (-1,-0.7) -- (1,-0.7) -- (1.5,-1.2) -- (2.0,-1.45) -- (2.3,-1.75) -- (2.3,-3) ;

\draw [thick] (-1,0.7) -- (-1,-0.7) ;
\draw [thick] (1,0.7) -- (1,-0.7) ;

\draw [thick] (1+0.5,0.7+0.5)--(-1-0.5,0.7+0.5);
\draw [thick] (1+0.5,-0.7-0.5)--(-1-0.5,-0.7-0.5);

\draw [thick] (2.0,1.45) -- (3.0,1.45);
\draw [thick] (-2.0,1.45) -- (-3.0,1.45) ;
\draw [thick] (2.0,-1.45) -- (3.0,-1.45) ;
\draw [thick]  (-2.0,-1.45) -- (-3.0,-1.45) ;

\draw [thick] (2.3,1.75) -- (3,1.75) ;
\draw [thick] (-2.3,1.75) -- (-3,1.75) ;
\draw [thick] (2.3,-1.75) -- (3,-1.75) ;
\draw [thick] (-2.3,-1.75) -- (-3,-1.75) ;

\draw [thick,,red!100] (-1,0) -- (1,0) ;
\draw [thick,,red!100] (-2.3,2.7) -- (2.3,2.7) ;
\draw [thick,,red!100] (-2.3,-2.7) -- (2.3,-2.7) ;

\node [red!100] at (0.9,0.3) {$\frac{1}{2}$ D1};
\node [red!100] at (0.9,3) {D1};
\node at (0,-0.5) {O7-4 D7};

\draw [thick] [->] [red!100] (0,2.8) --(0,3.2); 
\draw [thick] [->] [red!100] (0,-2.8) --(0,-3.2);

\end{tikzpicture}
\caption{type IIB brane diagram for the 5d $\mathcal{N}=1$ $Sp(2)$ gauge theory with $N_f=8$ hypermultiplets.}
\label{5dbrane2}
\end{figure}

Non-perturbative effect of the 5d gauge theory is essential for the duality. We first consider the general 5d $\mathcal{N}=1$ $Sp(N+1)$ gauge theories with $N_f=2N+8$ fundamental hypermultiplets. Type IIB brane diagram for $N=1$ case is given in Figure~\ref{5dbrane}. Instantons are realized by the D1 branes living on the D5 branes.
One should carefully use the string theory engineered ADHM construction. It contains unwanted extra degrees of freedoms \cite{Hwang:2014uwa}. For example, Figure~\ref{5dbrane2} shows the brane diagram for $Sp(N+1)$ gauge theory with $N_f=2N+6$ matters at $N=1$, which was considered in \cite{Seiberg:1996bd}. In this case D1 branes which can escape to infinity provide extra degrees of freedom. Their contribution to the instanton partition function can be computed separately. To obtain correct instanton partition function, one should subtract this extra contribution from the  ADHM quantum mechanical index. However, for 5d $Sp(N+1)$ gauge theory with $N_f=2N+8$ matters, we don't know how to identify the contribution of the extra degrees of freedom to the index.\footnote{$N_f=2N+6\,,2N+7$ cases are considered in \cite{Bergman:2015dpa}.} The extra states are supposed to be provided by the D1 branes moving vertically away from the D5 branes. We currently do not have technical controls of such extra states.\\

However one-instanton sector is special, because this sector is realized by the half-D1 brane stuck to O7$^-$ plane. The half-D1 brane can not escape to infinity, so it does not contains any extra degrees. For this reason, we expect that one can study the one-instanton sector of the general 5d $Sp(N+1)$ gauge theories with $N_f=2N+8$ fundamental hypermultiplets using the ADHM description.\\

5d index has the perturbative part and the instanton part $ Z^{\textrm{5d}} = Z^{\textrm{5d}}_{\textrm{pert}} Z^{\textrm{5d}}_{\textrm{inst}}$. The 5d Instanton partition functions for the $Sp(N+1)$ gauge group with matters are well-studied in \cite{Kim:2012gu,Hwang:2014uwa}. As we explained above, naive instanton partition functions can contain unwanted degrees freedom, so one should subtract this factor
\begin{align}
Z_{\textrm{inst}}=\frac{Z_{\textrm{ADHM}}}{Z_{\textrm{extra}}} =1+\sum_{k=1}^{\infty}q^k Z_k^{\textrm{5d},Sp(N+1)} \,,
\end{align}
$q$ is instanton fugacity and $k$ is instanton number. There is no $Z_{\textrm{extra}}$ factor for the one-instanton sector. $Z_k^{\textrm{5d},Sp(N+1)}$ is given by
\begin{align}
Z_k^{\textrm{5d},Sp(N+1)} = \textrm{Tr} \left[ (-1)^F e^{-\beta \{Q,Q^{\dagger}\}} e^{-\epsilon_1 (J_1+J_R)} e^{-\epsilon_2 (J_2 + J_R)} e^{-\alpha_i G_i} e^{-m_l F_l}\right] \,.
\label{ADHM}
\end{align}
$Q,Q^{\dagger}$ are two of the (0,4) supercharges of the ADHM QM system \cite{Hwang:2014uwa}. $J_1$ and $J_2$ are the Cartan generators of $SO(4)$ which rotating the $\mathbb{R}^4$. $J_R$ is the Cartan of $SU(2)_R$ R-symmetry. $G_i$ and $F_l$ are Cartans of $Sp(N+1)$ gauge group and $SO(4N+16)$ flavor symmetry group, and their conjugate chemical potentials are $\alpha_i$ and $m_l$. We will use the following fugacities convention $t = e^{-\epsilon_+},\; u = e^{-\epsilon_-},\; v_i= e^{-\alpha_i}$ and $y_l = e^{-m_l}$. \\

$Z_k^{\textrm{5d},Sp(N+1)}$ is given by the sum of $Z_k^{\pm}$, because the dual ADHM gauge group $O(k)$ has two disconnected sectors $O(k)^{\pm}$ \footnote{Actually $Sp(N+1)$ gauge theory has $\mathbb{Z}_2$ valued $\theta$ angle because $\pi_4(Sp(N+1)) = \mathbb{Z}_2$, so its index is given by 
\begin{displaymath}
Z_k^{\textrm{5d},Sp(N+1)} = \left\{ 
 \begin{array}{ll} 
 \frac{1}{2} (Z_k^+ + Z_k^-) & \,, \; \theta=0 \\
  \frac{(-1)^k}{2} (Z_k^+ - Z_k^-) & \,, \; \theta=\pi
 \end{array} \right. \;.
\end{displaymath}
But in our case, $\theta$ is not important. Its effect can be absorbed by redefinition of the flavor chemical potential.}
\begin{align}
Z_k^{\textrm{5d},Sp(N+1)} = \frac{1}{2} (Z_k^+ + Z_k^-) & \,.
\label{5dinst}
\end{align}
If one set $k=2n+\chi$ where $\chi=0$ or 1, $Z_k^{\pm}$ is given by
 \begin{align}
 Z_k^{\pm} = \frac{1}{|W|} \oint \prod_{I=1}^{n} \frac{d\phi_I}{2\pi i}Z_{\textrm{vec}}^{\pm}(\phi,\alpha_j ; \epsilon_{1,2}) \prod_{l} Z_{R_l}^{\pm}(\phi, \alpha_j, m_l ; \epsilon_{1,2}) \,,
 \end{align}
where Weyl factor $|W|$ is given by
\begin{align}
|W|^{\chi=0}_+ = 2^{n-1} n! \,, \;|W|^{\chi=1}_+ = 2^{n} n! \,, \; |W|^{\chi=0}_- = 2^{n-1} (n-1)! \,, \; |W|^{\chi=1}_- = 2^{n} n! \,.
\end{align}
$R_l$ denotes the representation of hypermultipet matters. See \cite{Hwang:2014uwa} for the details.

Vector multiplet part for $O(k)^{+}$ sector is given by
\begin{align}
Z_\textrm{vec}^{+} 
&= \left[  
\frac{1}{ 2\sinh\frac{\pm\epsilon_- + \epsilon_+}{2} \prod_{i=1}^{N+1} 2\sinh\frac{\pm\alpha_i +\epsilon_+}{2}} \prod_{I=1}^{n} \frac{2\sinh\frac{\pm\phi_I}{2} 2\sinh\frac{\pm\phi_I + 2 \epsilon_+}{2} }{2\sinh\frac{\pm\phi_I \pm\epsilon_-+\epsilon_+}{2}}
\right]^{\chi} \nonumber \\
& \quad \times \prod_{I=1}^{n} \frac{2\sinh\epsilon_+}{2\sinh\frac{\pm\epsilon_-+\epsilon_+}{2} \prod_{i=1}^{N+1} 2\sinh\frac{\pm\phi_I \pm\alpha_i +\epsilon_+}{2}}
\frac{\prod_{I>J}^n 2\sinh\frac{\pm\phi_I\pm\phi_J}{2} 2\sinh\frac{\pm\phi_I\pm\phi_J+2\epsilon_+}{2}}{\prod_{I=1}^n 2\sinh\frac{\pm2\phi_I\pm\epsilon_-+\epsilon_+}{2} \prod_{I>J}^n 2\sinh\frac{\pm\phi_I\pm\phi_J\pm\epsilon_-+\epsilon_+}{2}} \,,
\end{align}
and for $O(k)^{-}$ sector is given by
\begin{align}
Z_\textrm{vec}^{-} 
&=  
\frac{1}{ 2\sinh\frac{\pm\epsilon_- + \epsilon_+}{2} \prod_{i=1}^{N+1} 2\cosh\frac{\pm\alpha_i +\epsilon_+}{2}} \prod_{I=1}^{n} \frac{2\cosh\frac{\pm\phi_I}{2} 2\cosh\frac{\pm\phi_I + 2 \epsilon_+}{2} }{2\cosh\frac{\pm\phi_I \pm\epsilon_-+\epsilon_+}{2}}
 \nonumber \\
& \quad \times \prod_{I=1}^{n} \frac{2\sinh\epsilon_+}{2\sinh\frac{\pm\epsilon_-+\epsilon_+}{2} \prod_{i=1}^{N+1} 2\sinh\frac{\pm\phi_I \pm\alpha_i +\epsilon_+}{2}}
\frac{\prod_{I>J}^n 2\sinh\frac{\pm\phi_I\pm\phi_J}{2} 2\sinh\frac{\pm\phi_I\pm\phi_J+2\epsilon_+}{2}}{\prod_{I=1}^n 2\sinh\frac{\pm2\phi_I\pm\epsilon_-+\epsilon_+}{2} \prod_{I>J}^n 2\sinh\frac{\pm\phi_I\pm\phi_J\pm\epsilon_-+\epsilon_+}{2}} \,,
\end{align}
for $\chi=1$ and
\begin{align}
Z_\textrm{vec}^{-} 
&=  
\frac{2\cosh \epsilon_+}{ 2\sinh\frac{\pm\epsilon_- + \epsilon_+}{2} 2\sinh(\pm \epsilon_-+\epsilon_+) \prod_{i=1}^{N+1} 2\sinh(\pm\alpha_i +\epsilon_+)} \prod_{I=1}^{n-1} \frac{ 2\sinh(\pm\phi_I) 2\sinh(\pm\phi_I + 2 \epsilon_+) }{2\sinh(\pm\phi_I \pm\epsilon_-+\epsilon_+)}
 \nonumber \\
& \quad \times \prod_{I=1}^{n} \frac{2\sinh\epsilon_+}{2\sinh\frac{\pm\epsilon_-+\epsilon_+}{2} \prod_{i=1}^{N+1} 2\sinh\frac{\pm\phi_I \pm\alpha_i +\epsilon_+}{2}}
\frac{\prod_{I>J}^n 2\sinh\frac{\pm\phi_I\pm\phi_J}{2} 2\sinh\frac{\pm\phi_I\pm\phi_J+2\epsilon_+}{2}}{\prod_{I=1}^n 2\sinh\frac{\pm2\phi_I\pm\epsilon_-+\epsilon_+}{2} \prod_{I>J}^n 2\sinh\frac{\pm\phi_I\pm\phi_J\pm\epsilon_-+\epsilon_+}{2}} \,,
\end{align}
for $\chi=0$. Here and below, repeated $\pm$ signs in the argument of the $\sinh$ functions mean multiplying all such functions. 
For instance, \begin{align}
2\sinh(\pm a \pm b+c) \equiv 2\sinh(a+b+c)2\sinh(a-b+c)2\sinh(-a+b+c)2\sinh(-a-b+c) \,.
\end{align} \\
Fundamental hypermultiplets index contribution for $O(k)^{+}$ sector is given by
\begin{align}
Z_{\textrm{fund}}^{+} = \left(2\sinh\frac{m}{2} \right) \prod_{I=1}^n 2\sinh\frac{\pm\phi_I+m}{2} \,,
\end{align}
and for $O(k)^{-}$ sector is given by
\begin{align}
Z_{\textrm{fund}}^{-} =  2\cosh\frac{m}{2} \prod_{I=1}^n 2\sinh\frac{\pm\phi_I+m}{2} \,,
\end{align}
for $\chi=1$, and
\begin{align}
Z_{\textrm{fund}}^{-} = 2\sinh\frac{m}{2}\prod_{I=1}^{n-1} 2\sinh\frac{\pm\phi_I+m}{2} \,,
\end{align}
for $\chi=0$.\\

\paragraph{One-instanton: $N=0$} One can see that there is no contour integral for one-instanton sector. The one-instanton partition function for $Sp(1)$ gauge group with 8 fundamental matters is given by the sum of $Z_{1}^{\pm}$
\begin{align}
Z^{\textrm{5d},Sp(1)}_{k=1} & = \frac{1}{2} \left( \frac{1}{ 2\sinh\frac{\pm\epsilon_- + \epsilon_+}{2} 2\sinh\frac{\pm\alpha+\epsilon_+}{2}} \prod_{l=1}^{8} 2\sinh\frac{m_l}{2}
  +  \frac{1}{ 2\sinh\frac{\pm\epsilon_- + \epsilon_+}{2} 2\cosh\frac{\pm\alpha+\epsilon_+}{2}} \prod_{l=1}^{8} 2\cosh\frac{m_l}{2} \right) \nonumber \\
&= \frac{t}{(1-tu)(1-t/u)}\frac{t^2}{(1-t^2v^2)(1-t^2/v^2)} \Big[
(t+\frac{1}{t})\chi_{128}^{SO(16)}  - (v+\frac{1}{v}) \chi_{\overline{128}}^{SO(16)}
\Big] \,.
\label{5dinstresult}
\end{align}
It shows $SO(16)$ global symmetry.

\paragraph{Perturbative part: $N=0$} To obtain full 6d degrees of freedom, we must include the perturebative partition function
\begin{align}
Z^{\textrm{5d},Sp(1)}_{\textrm{pert}}
&= \textrm{PE}[ \frac{t}{(1-t u)(1-t/u)}\left(-(t+\frac{1}{t}) \chi^{Sp(1)}_{\textrm{adj},+}+ \chi^{Sp(1)}_{\textrm{fund},+} \chi_{16}^{SO(16)}(y_i)\right)] \nonumber \\
&= \textrm{PE}[ \frac{t}{(1-t u)(1-t/u)}\left(-(t+\frac{1}{t})v^2 + v \chi_{16}^{SO(16)}(y_i)\right)] \,,
\label{5dpertresult}
\end{align}
where $\chi^{Sp(1)}_{\textrm{R},+}$ denotes the $Sp(1)$ character of the representation R, but only sums over positive weights. This is because our index acquires contribution only from quarks, W-bosons, and their superpartners, but not from anti-quarks or anti-W-bosons. We will use this notation throughout the paper. 
Plethystic exponetial of $f(x)$ is defined by
\begin{align}
\textrm{PE}[f(x)] \equiv \exp\left( \sum_{n=1}^{\infty} \frac{1}{n} f(x^n)\right)
\label{pe}
\end{align}
where $x$ collectively denotes all the fugacities.
If we expand the 5d index $Z^{\textrm{5d}}=Z_{\textrm{pert}}^{\textrm{5d}}Z_{\textrm{inst}}^{\textrm{5d}}$ in terms of $Sp(1)$ fugacity $v$, it is exactly same as the E-strings elliptic genera in the sense of double expansion of the instanton fugacity $q$ and the string winding fugacity $w$ \footnote{In \cite{Kim:2014dza}, they exactly showed 5d-6d relation up to five-instantons and two-strings order.}. The 5d Coulomb vev fugacity $v$ is identified the 6d string winding number fugacity $w$, and the instanton fugacity $q$ becomes the string momentum fugacity $q$. Keeping in mind the 5d-6d fugacity relations and the $E_8$ Wilson line effect, we will study in the next two sections the 6d $Sp(N)$ gauge theories and their 5d $Sp(N+1)$ gauge theory descriptions. 
\section{6d SCFT with $Sp(1)$ gauge symmetry}
\label{sec:rank1}
In this section, we will study the circle compactified 6d SCFT with $Sp(1)$ gauge symmetry and its 5d $Sp(2)$ gauge theory description. Both theories have $N_f=10$ fundamental hypermultiplets. We confirm the duality by comparing the 5d instanton partition function and the elliptic genera of the self-dual strings in the 6d theory. The elliptic genera for the 6d $Sp(1)$ gauge theory are partially studied in \cite{Kim:2015fxa}. Main difference between \cite{Kim:2015fxa} and our computation is the presence of the $E_8$ Wilson line. The 6d theory can be Higgs to the E-string theory. So for duality to hold, one has to turn on the background $SO(20)$ Wilson line which reduces to the $E_8$ Wilson line after Higgsing. A natural guess is that the $SO(20)$ Wilson line will induce a shift $y_8\rightarrow y_8q^{-2}$, while leaving other $y_l$ unchanged. Indeed, we will show that the 5d and 6d indices agree with each other after this shift.
 \subsection{6d index}
To study full structure of the 6d index, we include not only the instanton soliton strings(or self-dual strings) part but also the 6d perturbative part $Z^{\textrm{6d}} = Z^{\textrm{6d}}_{\textrm{pert}}Z^{\textrm{6d}}_{\textrm{s.d}}$. The elliptic genus for self-dual strings is given by
\begin{align}
Z^{\textrm{6d},Sp(1)}_{\textrm{s.d}} = 1 + \sum_{n=1}^{\infty} w^n Z_n^{\textrm{6d},Sp(1)}\,,
\end{align}
where $Z_n^{\textrm{6d},Sp(1)}$ is given in \eqref{ellipticg}.
The matter contents of the 2d gauge theory description for the self-dual strings are given in Figure~\ref{2dquiver}. We are considering $N=1$ case, so there is an additional fundamental hypermultiplet contribution compared to the E-string theory.  To compare the 6d index with the 5d index, we will study the $q$-expanded form of the elliptic genera finally.
\paragraph{One-string} 
 One-string elliptic genus is similar with E-string case
\begin{align}
Z_{n=1}^{\textrm{6d},Sp(1)}
&= \frac{1}{2}\left( -Z_{1,[1]} +Z_{1,[2]} +Z_{1,[3]} -Z_{1,[4]} \right) \,,
\end{align}
where $Z_{1,[I]}$ are given by
\begin{align}
Z_{1,[I]} = 
- \frac{\eta^2}{\theta_1(\epsilon_1) \theta_1(\epsilon_2)} \cdot \prod_{l=1}^{10} \frac{\theta_I(m_l)}{\eta} \cdot \frac{\eta^2}{\theta_I(\epsilon_+ \pm \alpha)}  \,,
\end{align}
again after redefining the string winding fugacity $w \rightarrow wqy_8^{-1}$.
The $q$ expansion of this index is given by
\begin{align}
Z_{n=1}^{\textrm{6d},Sp(1)}& = q^0 \frac{t}{(1-tu)(1-t/u)}\left( \chi^{SO(20)}_{20}(y_i) -(v+\frac{1}{v})(t+\frac{1}{t}) \right) \nonumber \\
&\quad + q^1 \frac{t}{(1-tu)(1-t/u)}\frac{t^2}{(1-t^2v^2)(1-t^2/v^2)} \left( (t+\frac{1}{t}) \chi^{SO(20)}_{\overline{512}}(y_i) - (v+\frac{1}{v}) \chi^{SO(20)}_{512}(y_i) \right) +\mathcal{O}(q^2) \nonumber \\
& \equiv q^0 f_1(t,u,v,y_i) + q^1 \; Z_{1}^{\textrm{inst}}+\mathcal{O}(q^2) \,,
\label{6dsp11}
\end{align}
where $f_1$ and $Z_{1}^{\textrm{inst}}$ are defined by
\begin{align}
f_1(t,u,v,y_i) &= \frac{t}{(1-tu)(1-t/u)}\left( \chi^{SO(20)}_{20}(y_i) -(v+\frac{1}{v})(t+\frac{1}{t}) \right) \,, \\
Z_{1}^{\textrm{inst}} &=  \frac{t}{(1-tu)(1-t/u)}\frac{t^2}{(1-t^2v^2)(1-t^2/v^2)} \left( (t+\frac{1}{t}) \chi^{SO(20)}_{\overline{512}}(y_i) - (v+\frac{1}{v}) \chi^{SO(20)}_{512}(y_i) \right) \,.
\end{align}

\paragraph{Two-strings} Two-string elliptic genus is given by the sum of 7 discrete sectors
\begin{align}
\label{2wz1}
Z_{n=2}^{\textrm{6d},Sp(1)} = \frac{1}{2}Z_{2,[0]} +\frac{1}{4}\left( Z_{2,[1]} +Z_{2,[2]} +Z_{2,[3]} +Z_{2,[4]} +Z_{2,[5]} +Z_{2,[6]} \right) \,
\end{align}
where $Z_{2,[I]}$ are given by
\begin{align}
Z_{2,[0]} &= \oint \eta^2 du \frac{\theta_1(2\epsilon_+)}{i \eta} \cdot \frac{\eta^6}{\theta_1(\epsilon_1)\theta_1(\epsilon_2)\theta_1(\epsilon_1 \pm 2u)\theta_1(\epsilon_2 \pm 2u)} 
 \cdot \prod_{l=1}^{10} \frac{\theta_1(m_l \pm u)}{\eta} \cdot \frac{\eta^4}{\theta_1(\epsilon_+ \pm \alpha \pm u)} \,, \nonumber\\
Z_{2,[I]} &= \frac{\theta_1(a_v) \theta_1( 2\epsilon_+ +a_v)}{\eta^2} \cdot \frac{\eta^6}{\theta_1(\epsilon_1 +a_v)\theta_1(\epsilon_2+a_v)\theta_1(\epsilon_1+2a_{\pm})\theta_1(\epsilon_2+2a_{\pm})} \nonumber \\
&\quad \cdot \prod_{l=1}^{10} \frac{\theta_1(m_l+a_+) \theta_1(m_l+a_-)}{\eta^2} \cdot \frac{\eta^4}{\theta_1(\epsilon_+ \pm \alpha +a_+)\theta_1(\epsilon_+ \pm \alpha +a_-)}  \,, \; \textrm{for} \; I=1, \dots, 6.
\end{align}
Here $a_+,a_-,a_v(=a_++a_-)$ are given for $I=1,\dots,6$ by
\begin{align}
[I=1] : (a_+,a_-) &= (0,\frac{1}{2}) \,, &  [I=2] &: (a_+,a_-) =(\frac{\tau}{2},\frac{1+\tau}{2}) \,, \nonumber \\  
[I=3] : (a_+,a_-) &= (0,\frac{\tau}{2}) \,, & [I=4] &: (a_+,a_-) =(\frac{1}{2},\frac{1+\tau}{2}) \,, \\
[I=5] : (a_+,a_-) &= (0,\frac{1+\tau}{2}) \,, & [I=6] &: (a_+,a_-) =(\frac{1}{2},\frac{\tau}{2}) \,. \nonumber
\end{align}
$Z_{2,[0]}$ has a contour integral given by JK-residue \cite{Benini:2013nda,Benini:2013xpa}. The JK-residue prescription requires to sum over the residues at $u=-\frac{\epsilon_{1,2}}{2},\; -\frac{\epsilon_{1,2}}{2}+\frac{1}{2},\; -\frac{\epsilon_{1,2}}{2}+\frac{\tau}{2},-\frac{\epsilon_{1,2}}{2}+\frac{1+\tau}{2}$ from the symmetric and $u=-\epsilon_+ \pm \alpha$ from the fundamental hypermultiplet. The $SO(20)$ Wilson line shift changes the sign of $Z_{2,[I=1,2,5,6]}$
\begin{align}
\label{6dsp12}
Z_{n=1}^{\textrm{6d},Sp(1)}= \frac{1}{2}Z_{2,[0]} +\frac{1}{4}\left( -Z_{2,[1]} -Z_{2,[2]} +Z_{2,[3]} +Z_{2,[4]} -Z_{2,[5]} -Z_{2,[6]} \right) \,,
\end{align}
again after redefining the string winding fugacity $w \rightarrow wqy_8^{-1}$.
 $Z_{2,[I]}$ are obtained by
\begin{align}
Z_{2,[0]} 
&= \frac{1}{2}\frac{1}{\eta^{12}\theta_1(\epsilon_1)\theta_1(\epsilon_2)} \Bigg[ \sum_{i=1}^{4}\left( \frac{\prod_{l=1}^{10}\theta_i(m_l \pm\frac{\epsilon_1}{2})}{\theta_1(2\epsilon_1)\theta_1(\epsilon_2-\epsilon_1) \theta_i(\epsilon_+ \pm\alpha \pm\frac{\epsilon_1}{2} )}+(\epsilon_1 \rightarrow \epsilon_2)\right)\nonumber \\
&+ \left( \frac{\prod_{l=1}^{10} \theta_1(m_l \pm (\epsilon_+ +\alpha))}{\theta_1(\epsilon_1 \pm(\epsilon_+ + \alpha))\theta_1(\epsilon_2 \pm (\epsilon_+ + \alpha)) \theta_1(-2\alpha)\theta_1(2\epsilon_++2\alpha)} +(\alpha \rightarrow -\alpha)\right) \Bigg] \,,
\end{align}
\begin{align}
Z_{2,[1]} &= \frac{\theta_2(0)\theta_2(2\epsilon_+)\prod_{l=1}^{10}\theta_1(m_l)\theta_2(m_l)}{\eta^{12}\theta_1(\epsilon_1)^2\theta_1(\epsilon_2)^2 \theta_2(\epsilon_1)\theta_2(\epsilon_2)\theta_1(\epsilon_+\pm\alpha)\theta_2(\epsilon_+\pm\alpha)} \,, \; \nonumber \\
Z_{2,[2]} &= \frac{\theta_2(0)\theta_2(2\epsilon_+)\prod_{l=1}^{10}\theta_3(m_l)\theta_4(m_l)}{\eta^{12}\theta_1(\epsilon_1)^2\theta_1(\epsilon_2)^2 \theta_2(\epsilon_1)\theta_2(\epsilon_2)\theta_3(\epsilon_+\pm\alpha)\theta_4(\epsilon_+\pm\alpha)} \,, \; \nonumber \\
Z_{2,[3]} &= \frac{\theta_4(0)\theta_4(2\epsilon_+)\prod_{l=1}^{10}\theta_1(m_l)\theta_4(m_l)}{\eta^{12}\theta_1(\epsilon_1)^2\theta_1(\epsilon_2)^2 \theta_4(\epsilon_1)\theta_4(\epsilon_2)\theta_1(\epsilon_+\pm\alpha)\theta_4(\epsilon_+\pm\alpha)} \,, \; \nonumber \\
Z_{2,[4]} &= \frac{\theta_4(0)\theta_4(2\epsilon_+)\prod_{l=1}^{10}\theta_2(m_l)\theta_3(m_l)}{\eta^{12}\theta_1(\epsilon_1)^2\theta_1(\epsilon_2)^2 \theta_4(\epsilon_1)\theta_4(\epsilon_2)\theta_2(\epsilon_+\pm\alpha)\theta_3(\epsilon_+\pm\alpha)} \,, \; \nonumber \\
Z_{2,[5]} &= \frac{\theta_3(0)\theta_3(2\epsilon_+)\prod_{l=1}^{10}\theta_1(m_l)\theta_3(m_l)}{\eta^{12}\theta_1(\epsilon_1)^2\theta_1(\epsilon_2)^2 \theta_3(\epsilon_1)\theta_3(\epsilon_2)\theta_1(\epsilon_+\pm\alpha)\theta_3(\epsilon_+\pm\alpha)} \,, \; \nonumber \\
Z_{2,[6]} &= \frac{\theta_3(0)\theta_3(2\epsilon_+)\prod_{l=1}^{10}\theta_2(m_l)\theta_4(m_l)}{\eta^{12}\theta_1(\epsilon_1)^2\theta_1(\epsilon_2)^2 \theta_3(\epsilon_1)\theta_3(\epsilon_2)\theta_2(\epsilon_+\pm\alpha)\theta_4(\epsilon_+\pm\alpha)} \,. \; \nonumber \\
\end{align}
Finally $q$-expanded form of the two-strings elliptic genus \eqref{6dsp12} is 
\begin{align}
Z_{n=1}^{\textrm{6d},Sp(1)}
& = q^0\Bigg[ 
-\frac{t(t+\frac{1}{t})}{(1-tu)(1-t/u)} \nonumber \\\
&+\frac{1}{2} \left(\frac{t\left( \chi^{SO(20)}_{20}(y_i) -(v+\frac{1}{v})(t+\frac{1}{t}) \right)}{(1-tu)(1-t/u)}\right)^2 
+ \frac{1}{2}\left(\frac{t^2\left( \chi^{SO(20)}_{20}(y_i^2) -(v^2+\frac{1}{v^2})(t^2+\frac{1}{t^2}) \right)}{(1-t^2u^2)(1-t^2/u^2)}\right)
\Bigg] \nonumber \\
&+q^1 \Bigg[
\frac{t}{(1-tu)(1-t/u)}\frac{t^2}{(1-t^2v^2)(1-t^2/v^2)} \times 
\Bigg(
(t+\frac{1}{t})(v+\frac{1}{v}) \chi^{SO(20)}_{\overline{512}}(y_i) 
- (t+\frac{1}{t})^2 \chi^{SO(20)}_{512}(y_i) \nonumber \\
&+ \frac{t}{(1-tu)(1-t/u)}\left( \chi^{SO(20)}_{20}(y_i) -(v+\frac{1}{v})(t+\frac{1}{t}) \right) \times
 \left( (t+\frac{1}{t}) \chi^{SO(20)}_{\overline{512}}(y_i) - (v+\frac{1}{v}) \chi^{SO(20)}_{512}(y_i) \right)
\Bigg)
\Bigg]  \nonumber \\
&+\mathcal{O}(q^2) \\
& \equiv q^0 \left( f_2(t,u,v,y_i) + \frac{1}{2}\left( f_1(t,u,v,y_i) + f_1(t^2,u^2,v^2,y_i^2) \right)\right) + q^1 \left( Z_{2}^{\textrm{inst}} +  f_1(t,u,v,y_i) Z_{1}^{\textrm{inst}} \right) + \mathcal{O}(q^2) \,,
\end{align}
where $f_2(t,u,v,y_i)$ and $Z_{2}^{\textrm{inst}}$ are defined by
\begin{align}
f_2(t,u,v,y_i) &= -\frac{t(t+\frac{1}{t})}{(1-tu)(1-t/u)} \\
Z_{2}^{\textrm{inst}} &= \frac{t}{(1-tu)(1-t/u)}\frac{t^2}{(1-t^2v^2)(1-t^2/v^2)}
\left( (t+\frac{1}{t})(v+\frac{1}{v}) \chi^{SO(20)}_{\overline{512}}(y_i) - (t+\frac{1}{t})^2 \chi^{SO(20)}_{512}(y_i) \right) \,.
\end{align}
\paragraph{Three-strings} Three-string elliptic genus is given by the sum of 8 discrete sectors which are given by
\begin{align}
Z_{3,[1]} 
& =  -\oint \eta^2 du \frac{\theta_1(2\epsilon_+) \theta_1(2\epsilon_+\pm u) \theta_1(\pm u)}{i\eta^5}
 \cdot \frac{\eta^{12}}{\theta_1(\epsilon_{1,2})^2 \theta_1(\epsilon_{1,2} \pm u) \theta_1(\epsilon_{1,2} \pm 2u)} \nonumber \\
&\quad \cdot \prod_{l=1}^{10} \frac{\theta_1(m_l) \theta_1(m_l \pm u) }{\eta^3} 
\cdot \frac{\eta^6}{\theta_1(\epsilon_+\pm \alpha)\theta_1(\epsilon_+\pm \alpha \pm u)} \,, \\
Z_{3,[2]} 
& = - \oint \eta^2 du \frac{\theta_1(2\epsilon_+) \theta_2(2\epsilon_+\pm u) \theta_2(\pm u)}{i\eta^5}
 \cdot \frac{\eta^{12}}{\theta_1(\epsilon_{1,2})^2 \theta_2(\epsilon_{1,2} \pm u) \theta_1(\epsilon_{1,2} \pm 2u)} \nonumber \\
&\quad \cdot \prod_{l=1}^{10} \frac{\theta_2(m_l) \theta_1(m_l \pm u) }{\eta^3} 
\cdot \frac{\eta^6}{\theta_2(\epsilon_+\pm \alpha)\theta_1(\epsilon_+\pm \alpha \pm u)} \,, \\
Z_{3,[3]} 
& = - \oint \eta^2 du \frac{\theta_1(2\epsilon_+) \theta_3(2\epsilon_+\pm u) \theta_3(\pm u)}{i\eta^5}
 \cdot \frac{\eta^{12}}{\theta_1(\epsilon_{1,2})^2 \theta_3(\epsilon_{1,2} \pm u) \theta_1(\epsilon_{1,2} \pm 2u)} \nonumber \\
&\quad \cdot \prod_{l=1}^{10} \frac{\theta_3(m_l) \theta_1(m_l \pm u) }{\eta^3} 
\cdot \frac{\eta^6}{\theta_3(\epsilon_+\pm \alpha)\theta_1(\epsilon_+\pm \alpha \pm u)} \,, \\
Z_{3,[4]} 
& = - \oint \eta^2 du \frac{\theta_1(2\epsilon_+) \theta_4(2\epsilon_+\pm u) \theta_4(\pm u)}{i\eta^5}
 \cdot \frac{\eta^{12}}{\theta_1(\epsilon_{1,2})^2 \theta_4(\epsilon_{1,2} \pm u) \theta_1(\epsilon_{1,2} \pm 2u)} \nonumber \\
&\quad \cdot \prod_{l=1}^{10} \frac{\theta_4(m_l) \theta_1(m_l \pm u) }{\eta^3} 
\cdot \frac{\eta^6}{\theta_4(\epsilon_+\pm \alpha)\theta_1(\epsilon_+\pm \alpha \pm u)} \,, \\
Z_{3,[I']} 
& = - \frac{\theta_1(a_1+a_2)\theta_1(a_2+a_3)\theta_1(a_3+a_1) \theta_1(2\epsilon_++a_1+a_2)\theta_1(2\epsilon_++a_2+a_3)\theta_3(2\epsilon_++a_3+a_1)}{\eta^6} \nonumber \\
&\quad \cdot \frac{\eta^{12}}{\theta_1(\epsilon_{1,2}+2a_1)\theta_1(\epsilon_{1,2}+2a_2)\theta_1(\epsilon_{1,2}+2a_3) \theta_1(\epsilon_{1,2}+a_1+a_2)\theta_1(\epsilon_{1,2}+a_2+a_3)\theta_1(\epsilon_{1,2}+a_3+a_1)} \nonumber \\
&\quad \cdot \prod_{l=1}^{10} \frac{\theta_1(m_l+a_1)\theta_1(m_l+a_2)\theta_1(m_l+a_3)}{\eta^3} 
\cdot \frac{\eta^6}{\theta_1(\epsilon_+ \pm \alpha+ a_1)\theta_1(\epsilon_+ \pm \alpha+ a_2)\theta_1(\epsilon_+ \pm \alpha +a_3)} \,,
\end{align} 
where $a_1,a_2,a_3$ are given for $I'=1',2',3',4'$ by
\begin{align}
\label{2wdisc}
[I'=1'] &\rightarrow (a_1,a_2,a_3) = (\frac{1}{2},\frac{1+\tau}{2},\frac{\tau}{2}) \;\,, & [I'=2'] &\rightarrow (a_1,a_2,a_3) = (\frac{\tau}{2},\frac{1+\tau}{2},0) \;\,, \nonumber \\
[I'=3'] &\rightarrow (a_1,a_2,a_3) = (0,\frac{\tau}{2},\frac{1}{2}) \;\,, & [I'=4'] &\rightarrow (a_1,a_2,a_3) = (\frac{1}{2},\frac{1+\tau}{2},0) \;\,.
\end{align}
Each $Z_{3,[I]}$ has a contour integral. The non-zero JK-residues come from the poles at $u=-\frac{\epsilon_{1,2}}{2},\; -\frac{\epsilon_{1,2}}{2}+\frac{1}{2},\; -\frac{\epsilon_{1,2}}{2}+\frac{\tau}{2},-\frac{\epsilon_{1,2}}{2}+\frac{1+\tau}{2},-\epsilon \pm \alpha$ and $u= -\epsilon_{1,2} + \cdots$, where $\cdots$ part is decided by $\theta_i(\epsilon_{1,2}+u)=0$. After turning on the $SO(20)$ Wilson line, the three-string elliptic genus becomes
\begin{align}
Z_{n=3}^{\textrm{6d},Sp(1)}&= \frac{1}{4}\left( -Z_{3,[1]} + Z_{3,[2]} + Z_{3,[3]} - Z_{3,[4]} \right) +\frac{1}{8} \left( -Z_{3,[1']} + Z_{3,[2']} + Z_{3,[3']}- Z_{3,[4']}   \right) \nonumber \\
&= q^0 \left( \frac{1}{3}f_1(t^3,u^3,v^3,y_i^3) + \frac{1}{6}f_1(t,u,v,y_i)^3 +\frac{1}{2}f_1(t,u,v,y_i)f_1(t^2,u^2,v^2,y_u^2) + f_1(t,u,v,y_i)f_2(t,u,v,y_i)  \right) \nonumber \\
& + q \left( Z_{3}^{\textrm{inst}} + f_1(t,u,v,y_i)Z_{2}^{\textrm{inst}}+\left( f_2(t,u,v,y_i) + \frac{1}{2}\left( f_1(t,u,v,y_i) + f_1(t^2,u^2,v^2,y_i^2) \right)\right) Z_{1}^{\textrm{inst}}\right) +\mathcal{O}(q^2) \,,
\label{6dsp13}
\end{align}
where $Z_{3}^{\textrm{inst}}$ are defined by
\begin{align}
&Z_{3}^{\textrm{inst}}  \nonumber \\
&= \frac{t}{(1-tu)(1-t/u)}\frac{t^2}{(1-t^2v^2)(1-t^2/v^2)}
\left( (t+\frac{1}{t})(t^2+1+\frac{1}{t^2}) \chi^{SO(20)}_{\overline{512}}(y_i) - (v+\frac{1}{v})(t^2+1+\frac{1}{t^2}) \chi^{SO(20)}_{512}(y_i) \right)\,.
\end{align}
\paragraph{Perturbative index} The perturbative index of the theory on a circle is given by 
\begin{align}
\hspace*{-1.3cm}Z^{\textrm{6d},Sp(1)}_{\textrm{pert}} 
= \textrm{PE} \left[ 
\left( \frac{t}{(1-tu)(1-t/u)} \right) \left( -(t+\frac{1}{t}) \left( \chi^{Sp(1)}_{\textrm{adj},+} + \chi^{Sp(1)}_{\textrm{adj}} \frac{q^2}{1-q^2} \right) +
\left(  \chi^{Sp(1)}_{\textrm{fund},+} \chi^{SO(20)}_{\textrm{fund}}  +  \chi^{Sp(1)}_{\textrm{fund}} \chi^{SO(20)}_{\textrm{fund}}  \frac{q^2}{1-q^2} \right) \right)
\right] \nonumber \,,
\end{align}
where PE is defined in \eqref{pe}.
First term of the index comes from the 6d W-bosons and second term comes from the 6d fundamental quarks. The background $SO(20)$ Wilson line has no effect on the fields in the $SO(20)$ fundamental representation, and only affects spinor representation. So the perturbative index is unaffected by this Wilson line. In the exponent, we have only kept the contributions from BPS states with positive central charges in the regime $q \ll v \ll y_l^{\pm1}$.

\subsection{5d index}
Everything is same with $Sp(1)$ case. We only increase the gauge group rank by 1 and add two more fundamental hypermultiplets. So the index is generalization of \eqref{5dinstresult} and \eqref{5dpertresult}
\paragraph{Perturbative index}
\begin{align}
Z_{\textrm{pert}}^{\textrm{5d},Sp(2)}
&=\textrm{PE}\left[ \frac{t}{(1-tu)(1-t/u)}\left(-(t+\frac{1}{t}) \chi^{Sp(2)}_{\textrm{adj},+} + \chi^{Sp(2)}_{\textrm{fund},+} \chi^{SO(20)}_{\textrm{fund}} \right) \right] \nonumber \\
&=\textrm{PE}\Big[ \frac{t}{(1-tu)(1-t/u)}\Big(-(t+\frac{1}{t})
\left( v_1^2+v_2^2 +v_1  v_2 + \frac{v_1}{v_2}  \right) + \left( v_1+v_2 \right) \chi^{SO(20)}_{\textrm{fund}}(y_i) \Big) \Big] \,.
\label{5dsp2pert}
\end{align}
Here $v_i$ are defined below \eqref{ADHM}, and we chose our $Sp(2)$ positive roots by $2e_1,2e_2,e_1+e_2$ and $e_1-e_2$ where $e_1$ and $e_2$ are orthogonal unit vectors.  

\paragraph{One-instanton} One-instanton partition function is 
\begin{align}
Z^{\textrm{5d},Sp(2)}_{k=1} 
&= \frac{1}{2} \left( Z^{+}_{\textrm{vec}} Z^{+}_{\textrm{fund}}  +  Z^{-}_{\textrm{vec}}  Z^{-}_{\textrm{fund}}  \right) \nonumber \\
&= \frac{1}{2} \left( \frac{\prod_{j=1}^{10} 2\sinh\frac{m_j}{2}}{2\sinh\frac{\pm\epsilon_- + \epsilon_+}{2} \prod_{i=1}^{2} 2\sinh\frac{\pm \alpha_i + \epsilon_+}{2} } 
+ \frac{\prod_{j=1}^{10} 2\cosh\frac{m_j}{2}}{2\sinh\frac{\pm\epsilon_- + \epsilon_+}{2} \prod_{i=1}^{2} 2\cosh\frac{\pm \alpha_i + \epsilon_+}{2} } \right) \nonumber \\
&= \frac{t}{(1-tu)(1-t/u)}\frac{t^2}{(1-t^2v_1^2)(1-t^2/v_1^2)}\frac{t^2}{(1-t^2v_2^2)(1-t^2/v_2^2)} \nonumber \\
&\times \left( 
-\left( (v_1+\frac{1}{v_1})(v_2+\frac{1}{v_2}) + (t+\frac{1}{t})^2 \right) \chi^{SO(20)}_{512}(y_i) 
+\left( v_1+\frac{1}{v_1} + v_2+\frac{1}{v_2} \right)(t+\frac{1}{t}) \chi^{SO(20)}_{\overline{512}}(y_i)
\right) \,.
\label{5dsp2}
\end{align}
If we set $v_1=v,\;v_2=w$ and expand $Z^{\textrm{5d}}=Z^{\textrm{5d}}_{\textrm{pert}}Z^{\textrm{5d}}_{\textrm{inst}}$ in terms of $w$, it gives the same result as the 6d index $Z^{\textrm{6d}}$. Namely we checked that the $w$ expansion of $Z^{\textrm{5d}}$ completely argrees with \eqref{6dsp11},\eqref{6dsp12} and \eqref{6dsp13}.
\section{Generalization to the 6d SCFTs with $Sp(N)$ gauge group }
\label{sec:general}
In the previous section, we observed that the 6d string winding fugacity $w$ corresponds to one of the fugacities for the 5d $Sp(2)$ gauge symmetry in the instanton partition function. We can generalize this observation.
$Sp(N+1)$ group can be decomposed  into $Sp(1) \times Sp(N) \subset  Sp(N+1)$. We expect that the former $Sp(1) \sim SU(2)$ is responsible for the string winding fugacity, and the latter $Sp(N)$ gives the 6d gauge symmetry. We will confirm this assertion by comparing the 5d and 6d indices.
The 5d index for $Sp(N+1)$ gauge group and $N_f=2N+8$ fundamental hypermultiplets is given by
 \begin{align}
 \hspace*{-.5cm}Z^{\textrm{5d},Sp(N+1)}
 &=\textrm{PE}\left[ \frac{t}{(1-tu)(1-t/u)}\left(-(t+\frac{1}{t}) \chi^{Sp(N+1)}_{\textrm{adj},+} + \chi^{Sp(N+1)}_{\textrm{fund},+} \chi^{SO(4N+16)}_{\textrm{fund}} \right) \right] \nonumber \\
 &\quad  \times \left(1 + q\left( \frac{ \prod_{I=1}^{2N+8} 2\sinh\frac{m_I}{2} }{ 2\sinh\frac{\epsilon_1}{2} 2\sinh\frac{\epsilon_2}{2} \prod_{i=1}^{N+1} 2\sinh\frac{\epsilon_+ \pm \alpha_i}{2}}
 +\frac{ \prod_{I=1}^{2N+8} 2\cosh\frac{m_I}{2} }{ 2\cosh\frac{\epsilon_1}{2} 2\cosh\frac{\epsilon_2}{2} \prod_{i=1}^{N+1} 2\cosh\frac{\epsilon_+ \pm \alpha_i}{2}}   \right) +\mathcal{O}(q^2) \right) \,.
 \end{align}
 First line is the perturbative index and second line is the one-instanton partition function.
 To compare this result with the 6d index, we specially treat one of the Coulomb vev fugacity $v_{N+1}=e^{-\alpha_{N+1}} \equiv w$. Then $Sp(N+1)$ characters can be rewritten in terms of $Sp(N)$ characters and $w$
\begin{align}
\chi^{Sp(N+1)}_{\textrm{fund}}(v_i) 
& \equiv \sum_{i=1}^{N+1} \left( v_i + \frac{1}{v_i} \right) = \chi^{Sp(N)}_{\textrm{fund}}(v_i) +\left( w+\frac{1}{w} \right) \,, \\
\chi^{Sp(N+1)}_{\textrm{adj}}(v_i)
& \equiv \frac{ \left(\chi^{Sp(N+1)}_{\textrm{fund}}(v_i) \right)^2 + \chi^{Sp(N+1)}_{\textrm{fund}}(v_i^2) }{2} \nonumber \\
& = \frac{\left( \chi^{Sp(N)}_{\textrm{fund}}(v_i) +\left( w+\frac{1}{w} \right) \right)^2 + \chi^{Sp(N)}_{\textrm{fund}}(v_i^2) +\left( w^2+\frac{1}{w^2} \right)}{2} \nonumber \\
& = \chi^{Sp(N)}_{\textrm{adj}}(v_i) + \left(w+\frac{1}{w} \right) \chi^{Sp(N)}_{\textrm{fund}}(v_i) + w^2 + 1 + \frac{1}{w^2} \,.
\end{align}
Then the perturbative index becomes
\begin{align} 
Z_{\textrm{pert}}^{\textrm{5d},Sp(N+1)}
&= \textrm{PE}\left[ \frac{t}{(1-tu)(1-t/u)}\left(-(t+\frac{1}{t}) \chi^{Sp(N)}_{\textrm{adj},+} + \chi^{Sp(N)}_{\textrm{fund},+} \chi^{SO(4N+16)}_{\textrm{fund}} \right) \right] \nonumber \\
&\times \textrm{PE} \left[ \frac{t}{(1-tu)(1-t/u)}\left(-(t+\frac{1}{t}) w^2 + w \left( -(t+\frac{1}{t})\chi^{Sp(N)}_{\textrm{fund}} + \chi^{SO(4N+16)}_{\textrm{fund}} \right) \right) \right] \,,
\label{5dNpert}
\end{align}
where we only keep positive weights(roots) in the plethystic exponential. \\

We can expand the instanton partition function in terms of $w$
\begin{align} \label{5dNinst}
Z^{\textrm{5d},Sp(N+1)}_{k=1} &=\frac{1}{2} \Bigg( \frac{ \prod_{I=1}^{2N+8} 2\sinh\frac{m_I}{2} }{ 2\sinh\frac{\epsilon_1}{2} 2\sinh\frac{\epsilon_2}{2} \prod_{i=1}^{N} 2\sinh\frac{\epsilon_+ \pm \alpha_i}{2}} \frac{t}{(1-t w)(1-t/w)}  \nonumber \\
&\qquad\qquad + \frac{ \prod_{I=1}^{2N+8} 2\cosh\frac{m_I}{2} }{ 2\cosh\frac{\epsilon_1}{2} 2\cosh\frac{\epsilon_2}{2} \prod_{i=1}^{N} 2\cosh\frac{\epsilon_+ \pm \alpha_i}{2}}  \frac{t}{(1+t w)(1+t/w)} \Bigg) \nonumber \\
& = \frac{1}{2}w \left( -\frac{ \prod_{I=1}^{2N+8} 2\sinh\frac{m_I}{2} }{ 2\sinh\frac{\epsilon_1}{2} 2\sinh\frac{\epsilon_2}{2} \prod_{i=1}^{N} 2\sinh\frac{\epsilon_+ \pm \alpha_i}{2}}
+ \frac{ \prod_{I=1}^{2N+8} 2\cosh\frac{m_I}{2} }{ 2\cosh\frac{\epsilon_1}{2} 2\cosh\frac{\epsilon_2}{2} \prod_{i=1}^{N} 2\cosh\frac{\epsilon_+ \pm \alpha_i}{2}}  \right) \nonumber \\
&  - \frac{1}{2}w^2 \left( t + \frac{1}{t} \right) \left( \frac{ \prod_{I=1}^{2N+8} 2\sinh\frac{m_I}{2} }{ 2\sinh\frac{\epsilon_1}{2} 2\sinh\frac{\epsilon_2}{2} \prod_{i=1}^{N} 2\sinh\frac{\epsilon_+ \pm \alpha_i}{2}}
+ \frac{ \prod_{I=1}^{2N+8} 2\cosh\frac{m_I}{2} }{ 2\cosh\frac{\epsilon_1}{2} 2\cosh\frac{\epsilon_2}{2} \prod_{i=1}^{N} 2\cosh\frac{\epsilon_+ \pm \alpha_i}{2}}  \right) \nonumber \\
&  + \frac{1}{2}w^3 \left( t^2 +1+ \frac{1}{t^2} \right) \left( -\frac{ \prod_{I=1}^{2N+8} 2\sinh\frac{m_I}{2} }{ 2\sinh\frac{\epsilon_1}{2} 2\sinh\frac{\epsilon_2}{2} \prod_{i=1}^{N} 2\sinh\frac{\epsilon_+ \pm \alpha_i}{2}}
+ \frac{ \prod_{I=1}^{2N+8} 2\cosh\frac{m_I}{2} }{ 2\cosh\frac{\epsilon_1}{2} 2\cosh\frac{\epsilon_2}{2} \prod_{i=1}^{N} 2\cosh\frac{\epsilon_+ \pm \alpha_i}{2}}  \right)  \nonumber \\
& +\cdots \,.
 \end{align}
 Now we will compare this with the 6d index. Note that the first line of \eqref{5dNpert} is already same as the 6d perturbative index, so $w^0q^0$ orders clearly agree with each other. 
 \paragraph{One-string} Now we compare the 5d-6d results at $w^1q^0$ and $w^1q^1$ orders. One-string elliptic genus has following form
\begin{align}
Z_{n=1}^{\textrm{6d},Sp(N)} 
&= \frac{1}{2}\left( -Z_{1,[1]} +Z_{1,[2]} +Z_{1,[3]} -Z_{1,[4]} \right)  \,,
\label{6dN}
\end{align}
where $Z_{1,[I]}$ are given by
\begin{align}
Z_{1,[I]} = 
- \frac{\eta^2}{\theta_1(\epsilon_1) \theta_1(\epsilon_2)} \prod_{i=1}^{N}\frac{\eta^2}{\theta_I(\epsilon_+\pm \alpha_i)}\prod_{l=1}^{2N+8} \frac{\theta_I(m_l)}{\eta} \,.
\end{align}
After making $q$ expansion of $Z_{1,[I]}$, and after replacing all chemical potential by $z \rightarrow \frac{i z}{2\pi}$ (where $z$ denotes $\epsilon_{1,2} , \alpha_i, m_l$), one obtains
\begin{align}
Z_{1,[1]} & = \frac{ \prod_{l=1}^{2N+8} 2\sinh\frac{m_l}{2} }{ 2\sinh\frac{\epsilon_1}{2} 2\sinh\frac{\epsilon_2}{2} \prod_{i=1}^{N} 2\sinh\frac{\epsilon_+ \pm \alpha_i}{2}}q^1 + \mathcal{O}(q^2) \,,\\
Z_{1,[2]} & = \frac{ \prod_{l=1}^{2N+8} 2\cosh\frac{m_l}{2} }{ 2\cosh\frac{\epsilon_1}{2} 2\cosh\frac{\epsilon_2}{2} \prod_{i=1}^{N} 2\cosh\frac{\epsilon_+ \pm \alpha_i}{2}}q^1 + \mathcal{O}(q^2) \,, \\
Z_{1,[3]} & = \left( \frac{\sum_{l=1}^{2N+8}2\cosh\frac{m_l}{2} -\sum_{i=1}^{N} 2\cosh\frac{\epsilon_+\pm \alpha_i}{2}} {2 \cdot 2\sinh\frac{\epsilon_1}{2} 2\sinh\frac{\epsilon_1}{2}} \right)q^0+F(m_l,v_i,\epsilon_i) q^1  + \mathcal{O}(q^2) \,,\\
Z_{1,[4]} & = -\left( \frac{\sum_{l=l}^{2N+8}2\cosh\frac{m_l}{2} -\sum_{i=1}^{N} 2\cosh\frac{\epsilon_+\pm \alpha_i}{2}} {2 \cdot 2\sinh\frac{\epsilon_1}{2} 2\sinh\frac{\epsilon_1}{2}} \right)q^0+F(m_l,v_i,\epsilon_i) q^1  + \mathcal{O}(q^2) \,.
\end{align}
We do not write explicit form of $F(m_l,v_i,\epsilon_i)$ which is the coefficient of $q$ in $Z_{1,[3]}$ and $Z_{1,[4]}$, because they are canceled after summation. Then $w^1q^0$ term in \eqref{6dN} agrees with \eqref{5dNpert}. Also we have checked that $w^1q^1$ term agrees with the corresponding order of $Z^{5d}$.
\paragraph{Two-strings} We compare the 5d-6d results at $w^2q^0$ and $w^2q^1$ orders. Two-string elliptic genus is given by
\begin{align}
Z_{2,[0]} &= \oint \eta^2 du \frac{\theta_1(2\epsilon_+)}{i \eta} \cdot \frac{\eta^6}{\theta_1(\epsilon_1)\theta_1(\epsilon_2)\theta_1(\epsilon_1 \pm 2u)\theta_1(\epsilon_2 \pm 2u)} 
 \cdot \prod_{l=1}^{2N+8} \frac{\theta_1(m_l \pm u)}{\eta} \cdot \prod_{i=1}^{N}\frac{\eta^4}{\theta_1(\epsilon_+ \pm \alpha_i \pm u)} \,, \nonumber\\
Z_{2,[I]} &= \frac{\theta_1(a_v) \theta_1( 2\epsilon_+ +a_v)}{\eta^2} \cdot \frac{\eta^6}{\theta_1(\epsilon_1 +a_v)\theta_1(\epsilon_2+a_v)\theta_1(\epsilon_1+2a_{\pm})\theta_1(\epsilon_2+2a_{\pm})} \nonumber \\
&\quad \cdot \prod_{l=1}^{2N+8} \frac{\theta_1(m_l+a_+) \theta_1(m_l+a_-)}{\eta^2} \cdot \prod_{i=1}^{N}\frac{\eta^4}{\theta_1(\epsilon_+ \pm \alpha_i +a_+)\theta_1(\epsilon_+ \pm \alpha_i +a_-)}  \,,
\end{align}
where discrete sector $I$ is same as \eqref{2wdisc}. There are additional poles from symmetric hypermultiplets, which are given by $u_*=-\epsilon_+ \pm\alpha_i$ for all $i$. Now we can obtain general form of two-strings 
elliptic genus.
\begin{align}
Z_{2,[0]} 
&= \frac{1}{2}\frac{1}{\eta^{12}\theta_1(\epsilon_1)\theta_1(\epsilon_2)} \Bigg[ \sum_{i=1}^{4}\left( \frac{\prod_{l=1}^{2N+8}\theta_i(m_l \pm\frac{\epsilon_1}{2})}{\theta_1(2\epsilon_1)\theta_1(\epsilon_2-\epsilon_1)\prod_{m=1}^{N} \theta_i(\epsilon_+ \pm\alpha_m \pm\frac{\epsilon_1}{2} )}
+(\epsilon_1 \rightarrow \epsilon_2)\right)\nonumber \\
+ \sum_{n=1}^{N} & \Bigg( \frac{\prod_{l=1}^{2N=8} \theta_1(m_l \pm (\epsilon_+ +\alpha_n))}{\theta_1(\epsilon_1 \pm2(\epsilon_+ + \alpha_n))\theta_1(\epsilon_2 \pm 2(\epsilon_+ + \alpha_n)) \theta_1(-2\alpha_n)\theta_1(2\epsilon_++2\alpha_n)\prod_{\substack{m=1\\ m\neq n}}^N \theta_1(- \alpha_n \pm \alpha_m)\theta_1(2\epsilon_++\alpha_n \pm \alpha_m) } \nonumber \\
& \quad +(\alpha_n \rightarrow -\alpha_n) \Bigg) 
\Bigg] \,,
\end{align}
\begin{align}
Z_{2,[1]} &= \frac{\theta_2(0)\theta_2(2\epsilon_+)\prod_{l=1}^{2N+8}\theta_1(m_l)\theta_2(m_l)}{\eta^{12}\theta_1(\epsilon_1)^2\theta_1(\epsilon_2)^2 \theta_2(\epsilon_1)\theta_2(\epsilon_2) \prod_{m=1}^N \theta_1(\epsilon_+\pm\alpha_m)\theta_2(\epsilon_+\pm\alpha_m)} \,, \; \nonumber \\
Z_{2,[2]} &= \frac{\theta_2(0)\theta_2(2\epsilon_+)\prod_{l=1}^{2N+8}\theta_3(m_l)\theta_4(m_l)}{\eta^{12}\theta_1(\epsilon_1)^2\theta_1(\epsilon_2)^2 \theta_2(\epsilon_1)\theta_2(\epsilon_2) \prod_{m=1}^N \theta_3(\epsilon_+\pm\alpha_m)\theta_4(\epsilon_+\pm\alpha_m)} \,, \; \nonumber \\
Z_{2,[3]} &= \frac{\theta_4(0)\theta_4(2\epsilon_+)\prod_{l=1}^{2N+8}\theta_1(m_l)\theta_4(m_l)}{\eta^{12}\theta_1(\epsilon_1)^2\theta_1(\epsilon_2)^2 \theta_4(\epsilon_1)\theta_4(\epsilon_2) \prod_{m=1}^N \theta_1(\epsilon_+\pm\alpha_m)\theta_4(\epsilon_+\pm\alpha_m)} \,, \; \nonumber \\
Z_{2,[4]} &= \frac{\theta_4(0)\theta_4(2\epsilon_+)\prod_{l=1}^{2N+8}\theta_2(m_l)\theta_3(m_l)}{\eta^{12}\theta_1(\epsilon_1)^2\theta_1(\epsilon_2)^2 \theta_4(\epsilon_1)\theta_4(\epsilon_2) \prod_{m=1}^N \theta_2(\epsilon_+\pm\alpha_m)\theta_3(\epsilon_+\pm\alpha_m)} \,, \; \nonumber \\
Z_{2,[5]} &= \frac{\theta_3(0)\theta_3(2\epsilon_+)\prod_{l=1}^{2N+8}\theta_1(m_l)\theta_3(m_l)}{\eta^{12}\theta_1(\epsilon_1)^2\theta_1(\epsilon_2)^2 \theta_3(\epsilon_1)\theta_3(\epsilon_2) \prod_{m=1}^N \theta_1(\epsilon_+\pm\alpha_m)\theta_3(\epsilon_+\pm\alpha_m)} \,, \; \nonumber \\
Z_{2,[6]} &= \frac{\theta_3(0)\theta_3(2\epsilon_+)\prod_{l=1}^{2N+8}\theta_2(m_l)\theta_4(m_l)}{\eta^{12}\theta_1(\epsilon_1)^2\theta_1(\epsilon_2)^2 \theta_3(\epsilon_1)\theta_3(\epsilon_2) \prod_{m=1}^N \theta_2(\epsilon_+\pm\alpha_m)\theta_4(\epsilon_+\pm\alpha_m)} \,. \; \nonumber \\
\end{align}
After plugging these into the \eqref{6dsp12}, one can obtain two-string elliptic genus.
We compared the $q$-expanded form of this elliptic genus with 5d index by increasing $N$ up to $N=8$, and we saw perfect agreements of the two results. We also checked the agreement of three-strings elliptic genus up to $N=3$.
 \section{Conclusion}
 \label{sec:conc}
 We studied the 6d SCFTs compactified on a circle with $Sp(N)$ gauge symmetry and $N_f=2N+8$ fundamental hypermultiplets. In particular, we tested the 5d $Sp(N+1)$ gauge theory descriptions of the 6d theories. We compared the Witten indices of the 5d and 6d theories. For the 5d instanton partition function, the usual ADHM construction contains unwanted string theory degrees except in the one-instanton sector. We observe perfect agreement of the two indices in double expansion of the string winding fugacity $w$ and the instanton fugacity $q$ up to $w^3q^1$ order. As usual, the 5d instanton charge is mapped to the 6d KK momentum mode. The 5d $Sp(N+1)$ gauge group is decomposed into the $Sp(1) \times Sp(N)$, and the former $Sp(1)$ charge is mapped the 6d self-dual string winding number. The fugacities for the latter 5d $Sp(N)$ gauge symmetry and $SO(4N+16)$ flavor symmetry are mapped to the 6d $Sp(N)$ gauge symmetry and $SO(4N+16)$ flavor symmetry. We have also observed that the background $SO(4N+16)$ Wilson line plays crucial roles in these 5d-6d dualities, similar to the $E_8$ Wilson line in the E-string theory. These results provide the detailed rules of the dualities proposed by \cite{Hayashi:2015zka}. \\
 
   The natural question is what happens if we naively compute higher instanton partition functions using \eqref{5dinst}. Our naive computation shows disagreements with the result predicted by the ellipitic genus of self-dual strings. 
 Difference between two results must comes from the extra degrees in the string engineered ADHM construction. We hope this result gives the better understanding of the extra degrees in the brane system. \\
   
   We can also try to check the duality between 5d gauge theory with $Sp(2)$ and $SU(3)$ gauge groups, each having $10$ fundamental hypermulitplets\cite{Tachikawa:2015mha,Yonekura:2015ksa,Hayashi:2015fsa,Gaiotto:2015una}. The SU(3) gauge theory is also conjectured to uplift to the same 6d $Sp(1)$ SCFT on a circle. Although the string theory engineered ADHM construction of the $SU(3)$ gauge theories have extra degrees too, we can detour this problem by introducing anti-symmetric hypermultiplets. Anti-symmetric representation is same as (anti-)fundamental representation in $SU(3)$ group. So the 5d $SU(3)$ gauge theory with $10$ fundamental hypermultiplets can be regarded as the gauge theory with 8 fundamental and 2 anti-symmetric hypermultiplets. This trivial change of viewpoint affects the details of the ADHM construction. Such alternative ADHM descriptions are sometimes shown to provide more useful description of instantons \cite{Gaiotto:2015una}. It is interesting to see if their ideas apply to our system.
  
\vspace{0.8cm}

\noindent{\bf\large Acknowledgements}

\noindent
The author would like to thank Yoonseok Hwang, Ki-Hong Lee, and Sangmin Lee for helpful disccusions. I would especially like to thank Seok Kim for suggesting this project, valuable discussions, and careful reading of the manuscript. I also thank Sung-Soo Kim for informing me of the publication schedule of \cite{sskim}, which has an overlap with this paper. The work of YY is supported by Samsung Science and Technology Foundation under Project Number SSTF-BA1402-08 and BK 21 Plus Program. 
 
\pagebreak
\providecommand{\href}[2]{#2}\begingroup\raggedright
\endgroup

\end{document}